\documentclass{Interspeech2024}
\usepackage{multirow}
\usepackage{graphicx} 
\usepackage{verbatim}
\usepackage{cite}
\usepackage{makecell}

\interspeechcameraready

\title{SVSNet+: Enhancing Speaker Voice Similarity Assessment Models with Representations from Speech Foundation Models}

\name[affiliation={1,2}]{Chun}{Yin}
\name[affiliation={1}]{Tai-Shih}{Chi}
\name[affiliation={2}]{Yu}{Tsao}
\name[affiliation={2}]{and Hsin-Min}{Wang}

\address{
  $^1$National Yang Ming Chiao Tung University, Taiwan\\
  $^2$Academia Sinica, Taiwan}
\email{nash814@gmail.com, tschi@nycu.edu.tw, yu.tsao@citi.sinica.edu.tw, whm@iis.sinica.edu.tw}


\begin{document}

\maketitle

\begin{abstract}
Representations from pre-trained speech foundation models (SFMs) have shown impressive performance in many downstream tasks. However, the potential benefits of incorporating pre-trained SFM representations into speaker voice similarity assessment have not been thoroughly investigated. In this paper, we propose SVSNet+, a model that integrates pre-trained SFM representations to improve performance in assessing speaker voice similarity. Experimental results on the Voice Conversion Challenge 2018 and 2020 datasets show that SVSNet+ incorporating WavLM representations shows significant improvements compared to baseline models. In addition, while fine-tuning WavLM with a small dataset of the downstream task does not improve performance, using the same dataset to learn a weighted-sum representation of WavLM can substantially improve performance. Furthermore, when WavLM is replaced by other SFMs, SVSNet+ still outperforms the baseline models and exhibits strong generalization ability.
\end{abstract}

\section{Introduction}

A useful voice conversion (VC) system can be used in many areas, such as dubbing, personalized virtual assistants, preserving voices, and aiding in voice recovery after surgery. However, effectively evaluating the performance of such systems remains challenging. As a common evaluation for VC systems, the speaker voice similarity assessment aims to evaluate the resemblance between generated speech and natural speech. The evaluation can be objective \cite{das2020predictions} or subjective. While the results of subjective evaluation are closer to human perception, the time and cost required to conduct assessments such as listening tests are considerable. Therefore, automated and efficient methods for assessing speaker voice similarity are valuable.

In \cite{9716822}, Hu et al. proposed SVSNet, an end-to-end neural model for speaker voice similarity assessment, and showed satisfactory performance at both the utterance and system levels. SVSNet takes the raw speech waveforms as inputs and processes them with an encoder consisting of a SincNet module, four stacked residual-skipped-WaveNet convolution (rSWC) layers, and a BLSTM layer. However, given the constraint of limited training data, the encoder may not be sufficiently adept at extracting meaningful representations from the waveforms. Therefore, to further reinforce the speaker voice similarity assessment model, we propose SVSNet+, which integrates a large pre-trained speech foundation model (SFM) to extract speech representations. By leveraging a pre-trained SFM that learns effective speech representations from large-scale training data, SVSNet+ can acquire valuable information for the similarity prediction task. When evaluated on the Voice Conversion Challenge 2018 \cite{lorenzotrueba18_odyssey} and 2020 \cite{yi20_vccbc} (VCC2018 and VCC2020) datasets, SVSNet+ significantly outperforms previous work at the system level and demonstrates strong generalization ability. This study contributes to further future exploration of applying pre-trained SFM representations to speaker voice similarity assessment.

\section{Related Work}

\subsection{Pre-trained speech foundation models}

Pre-trained SFMs can undergo training in a supervised or unsupervised learning manner. Although supervised learning typically excels over unsupervised learning in various tasks, the process of collecting large-scale labeled data can be time-consuming and sometimes impractical. Therefore, many state-of-the-art pre-trained models are based on self-supervised learning (SSL), allowing them to acquire meaningful representations from large amounts of unlabeled data \cite{liu2022audio}. As one of the most commonly used models, wav2vec 2.0 \cite{NEURIPS2020_92d1e1eb} acquires contextual information by discerning representations that correspond to true quantized latent speech representations. On the other hand, HuBERT \cite{9585401} utilizes clustering to generate labels and predicts hidden cluster assignments for masked speech representations. Modified from HuBERT, WavLM \cite{9814838} uses a larger dataset during pre-training for joint learning of masked speech prediction and denoising. Moreover, the Massively Multilingual Speech (MMS) model \cite{pratap2023scaling} expands the number of languages in the training dataset and builds pre-trained models covering 1,406 languages based on wav2vec 2.0. Unlike the aforementioned SSL pre-trained models, Whisper \cite{radford2023robust} is an SFM trained in a weakly supervised manner, using a multitask training format, showcasing not only high robustness but also strong generalization ability. All of these SFMs have been employed to enhance performance on diverse speech processing tasks, demonstrating remarkable effectiveness. In this study, we employ all these SFMs to extract speech representations and evaluate their suitability for the speaker voice similarity assessment task.

\subsection{Speech assessment tasks using SFM-extracted representations}

In recent years, SFM representations have been applied in various tasks, such as speech enhancement (SE) \cite{hung22_interspeech}, automatic speech recognition (ASR) \cite{10096426}, automatic speaker verification (ASV) \cite{9747814}, and voice conversion (VC) \cite{chen23h_interspeech}. In speech assessment tasks, it is a prevalent practice to leverage pre-trained SSL representations for mean opinion score (MOS) prediction \cite{9746395, saeki22c_interspeech, tseng21b_interspeech, yang22o_interspeech}. The latent representations extracted by wav2vec 2.0, HuBERT, and WavLM have been proven beneficial for these tasks. Furthermore, recent work by Zezario et al. \cite{zezario2023study} also showed that Whisper and MMS representations help predict human-perceived speech quality and intelligibility.

\section{Proposed Method}

The architecture of SVSNet+ is shown in Fig. 1. The waveforms of the test and reference utterances \(X_T \) and \(X_R \) are fed to a pre-trained SFM, which encodes them into layer-wise representations. Next, the corresponding weighted-sum representations are derived from layer-wise representations, and a linear layer is used to adjust the representation dimension. Then, the representations \(R_T \) and \(R_R \) are aligned by the co-attention module in both directions for maintaining symmetry. Afterwards, the distance module calculates the distance between \(R_T \) and \(\hat{R}_R \) and that between \(R_R \) and \(\hat{R}_T \). Finally, the prediction module uses these two distances to calculate a similarity score.

\subsection{Pre-trained model and weighted sum}

In this study, we study several SFMs, each containing a feature extractor and a transformer encoder, as shown in Fig. 1(b). For SSL-based SFMs, the feature extractors are CNN-based encoders that generate feature sequences at a frame rate of 20ms for audio sampled at 16kHz. For Whisper, the feature extractor first preprocesses the 16kHz audio input into 30-second chunks by zero-padding or trimming. Each chunk is then transformed into an 80-channel log-magnitude Mel spectrogram at a frame rate of 10ms and further processed by an encoder consisting of two convolutional layers and a GELU activation function. The stride of the second convolutional layer is 2 \cite{radford2023robust}. Therefore, the down-sampling factor of the feature extractor in each pre-trained SFM is 320x. For all SFMs, the extracted features are fed to the transformer encoder and processed by $L$ hidden layers. Finally, to exploit the information from each hidden layer, the representations generated from all hidden layers are combined using the weighted sum module:
\begin{align}
  R_{WS} &:= \sum_{\ell=0}^{L-1} w^{\ell}R^{\ell},
  \label{equation:eq1}
\end{align}
where \(w^{\ell} \ge 0 \) is the learnable weight for layer $\ell$ and \({\textstyle \sum_{\ell}} w^{\ell} = 1 \), and \(R^{\ell} \) is the representation of layer $\ell$. The weighted-sum representation is then passed through an additional linear layer for dimension adjustment.

\subsection{Co-attention module}

Following \cite{9716822}, the co-attention module is used to align the representation of one input with that of the other input by
\begin{align}
  \hat{R}_R &= Attention(R_T,R_R,R_R), \nonumber \\
  \hat{R}_T &= Attention(R_R,R_T,R_T),
  \label{equation:eq2}
\end{align}
and output two aligned pairs \((R_T, \hat{R}_R) \) and \((R_R, \hat{R}_T) \), which will be input to the distance module. In this study, the scaled dot-product attention mechanism \cite{NIPS2017_3f5ee243} is implemented.

\subsection{Distance module and prediction module}

Following \cite{9716822}, the utterance embedding is obtained by averaging its representations over time, and the 1-norm distance of each dimension of two embeddings is calculated:
\begin{align}
  D_{T,R} &= \parallel Mean(R_T) - Mean(\hat{R}_R)\parallel_1, \nonumber \\
  D_{R,T} &= \parallel Mean(R_R) - Mean(\hat{R}_T)\parallel_1.
  \label{equation:eq3}
\end{align}
Then, the prediction module takes in the two distances to derive the similarity scores:
\begin{align}
  \hat{S}_T &= \sigma(f_{lin2}(\rho_{ReLU}(f_{lin1}(D_{T,R})))), \nonumber \\ 
  \hat{S}_R &= \sigma(f_{lin2}(\rho_{ReLU}(f_{lin1}(D_{R,T})))),
  \label{equation:eq4}
\end{align}
where \(\sigma(.) \) is an activation function, \(f_{lin1}(.) \) and \(f_{lin2}(.) \) are linear layers, and \(\rho_{ReLU}(.) \) is the rectified linear unit (ReLU) activation function. There are two types of prediction modules: regression and classification. For regression tasks, the identity function is used as the activation function, while for classification tasks, the softmax function is used. The output size of the second linear layer is 1 (for regression) and 4 (for classification). The final score is the average of the two predicted scores:
\begin{align} 
  \hat{S} &= (\hat{S}_T + \hat{S}_R)/2.
  \label{equation:eq5}
\end{align}
The training objective of the proposed model is to match these scores with the corresponding human-labeled similarity scores in the training set.

\begin{figure}[t]
  \centering
  \includegraphics[width=\linewidth]{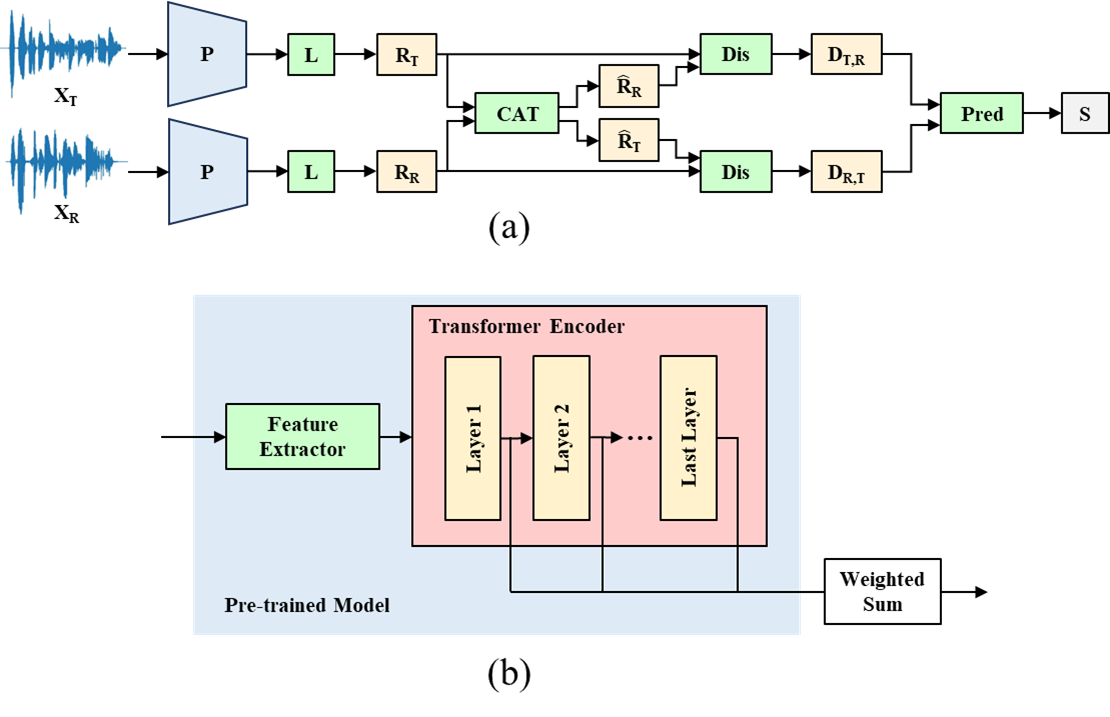}
  \vspace{-0.7cm}
  \caption{(a) The architecture of SVSNet+. P, L, CAT, Dis, and Pred respectively denote the pre-trained model, linear
layer, co-attention module, distance module, and prediction module. (b) The pre-trained model followed by the weighted sum module.}
  \label{fig:model_Architecture}
\end{figure}

\section{Experiments}

\subsection{Experimental setup}

\subsubsection{Datasets}

Following \cite{9716822}, the proposed method was evaluated on the Voice Conversion Challenge 2018 and 2020 (VCC2018 and VCC2020) datasets. In each challenge, participants submitted audio files produced by their VC systems. Subsequently, subjective listening tests were conducted to evaluate these systems. Subjects were asked to rate the converted utterances based on their quality and similarity to the reference utterances. This study focuses on similarity assessment.

In the VCC2018 dataset \cite{lorenzotrueba18_odyssey}, the utterances were derived from 36 VC systems and two reference systems. A total of 21,562 converted-natural utterance pairs were evaluated, yielding 30,864 speaker similarity scores ranging from 1 to 4. Higher scores indicate that the speakers in an utterance pair sound more similar to each other. we split the dataset into training and test sets, with 24,864 and 6,000 ratings, respectively.

In VCC2020 \cite{yi20_vccbc}, there were two tasks, intra-lingual semi-parallel VC and cross-lingual VC. Of the 33 participants, 31 teams submitted results for the intra-lingual task, and 28 teams submitted results for the cross-lingual task. There were 5,840 converted-target utterance pairs. Each pair was evaluated by multiple subjects, and the average score was used as the final score for the pair. Including source-target and target-target utterance pairs as lower and upper performance bounds, the full VCC2020 test set contains 6,090 scored pairs.

\subsubsection{Evaluation metrics}

Performance was evaluated in terms of linear correlation coefficient (LCC) \cite{24cdd36d-ce47-3cc8-a397-07b95efc6732}, Spearman’s rank correlation coefficient (SRCC) \cite{85f53594-1421-3dd6-8318-2f0686ede307}, and mean squared error (MSE) at the utterance and system levels. The utterance-level evaluation was calculated based on the predicted score and the human-labeled score for each test utterance pair, while the system-level evaluation was calculated based on the average predicted score and average human-labeled score for each system. System-level evaluation is more valuable because it directly ranks systems.

\subsubsection{Pre-trained models}

We evaluated several SFMs. WavLM-Large \cite{9814838}, wav2vec 2.0-Large (LV-60) \cite{NEURIPS2020_92d1e1eb}, MMS-300M, and MMS-1B \cite{pratap2023scaling} were obtained from their GitHub websites. HuBERT-Large and HuBERT-XLarge \cite{9585401} were accessed via the pipelines subpackage of \emph{TorchAudio} \cite{9747236, 10389648}. Whisper-Medium and Whisper-Large \cite{radford2023robust} were taken from HuggingFace's \emph{Transformers} library \cite{wolf-etal-2020-transformers}. Note that all SFMs are versions without any fine-tuning. Furthermore, Whisper's original settings involve preprocessing the input waveform into 30-second chunks via zero-padding or trimming. Since all other SFMs use full-length audio inputs and each utterance in the VCC2018 and VCC2020 datasets is shorter than 10 seconds, we modified Whisper's configuration for a fair comparison. Specifically, the chunk length was set to 10 seconds, and the resulting representation was trimmed back to the original length in the time domain.

\begin{table*}[t]
\begin{center}
\caption{Performance of different configurations of SVSNet+ with WavLM-Large evaluated on VCC2018. Modules included in SVSNet+ are marked after underline, where r, B, P, L, F, and W denote rSWC, BLSTM, pre-trained model, linear layer, fine-tuning, and weighted sum, respectively.}
\vspace{-0.2cm}
\scalebox{0.65}{
\begin{tabular}{lcccccccccccccc}
\toprule
\multirow{2}{*}{\textbf{Method}} & \multirow{2}{*}{\textbf{SincNet}} & \multirow{2}{*}{\textbf{rSWC}} & \multirow{2}{*}{\textbf{BLSTM}} & \multirow{2}{*}{\textbf{\makecell[c]{Pre-trained \\ Model}}} & \multirow{2}{*}{\textbf{\makecell[c]{Linear \\ Layer}}} & \multirow{2}{*}{\textbf{\makecell[c]{Fine- \\ tuning}}} & \multirow{2}{*}{\textbf{\makecell[c]{Pre-trained \\ Representation Type}}} & \multirow{2}{*}{\textbf{Model Type}} & \multicolumn{3}{c}{\textbf{Utterance-level}} & \multicolumn{3}{c}{\textbf{System-level}} \\ \cmidrule(r){10-12} \cmidrule(r){13-15}
 &  &  &  &  &  &  &  &  & \textbf{LCC ↑} & \textbf{SRCC ↑} & \textbf{MSE ↓} & \textbf{LCC ↑} & \textbf{SRCC ↑} & \textbf{MSE ↓} \\
\midrule
SVSNet(R) & O & O & O & X & X & X & - & Regression & 0.530 & 0.531 & 0.952 & 0.924 & 0.902 & 0.009 \\
SVSNet(C) & O & O & O & X & X & X & - & Classification & 0.540 & 0.542 & 1.160 & 0.914 & 0.874 & 0.032 \\
SVSNet+\_BP(R) & X & X & O & O & X & X & LL & Regression & 0.546 & 0.547 & 0.899 & 0.952 & 0.938 & 0.005 \\
SVSNet+\_BPL(R) & X & X & O & O & O & X & LL & Regression & 0.521 & 0.524 & 0.994 & 0.952 & 0.950 & 0.015 \\
SVSNet+\_BPL(C) & X & X & O & O & O & X & LL & Classification & 0.484 & 0.488 & 1.358 & 0.932 & 0.919 & 0.028 \\
SVSNet+\_rBPL(R) & X & O & O & O & O & X & LL & Regression & 0.523 & 0.524 & 0.966 & 0.926 & 0.894 & 0.008 \\
SVSNet+\_PL(R) & X & X & X & O & O & X & LL & Regression & 0.546 & 0.548 & 0.902 & 0.943 & 0.953 & 0.006 \\
SVSNet+\_PLF(R) & X & X & X & O & O & O & LL & Regression & 0.498 & 0.502 & 1.022 & 0.941 & 0.923 & 0.007 \\
SVSNet+\_PLW(R) & X & X & X & O & O & X & WS & Regression & 0.567 & 0.567 & 0.879 & 0.953 & 0.957 & \textbf{0.005} \\
SVSNet+\_PLFW(R) & X & X & X & O & O & O & WS & Regression & 0.537 & 0.540 & 0.921 & 0.937 & 0.927 & 0.009 \\
SVSNet+\_P(R) & X & X & X & O & X & X & LL & Regression & 0.506 & 0.501 & 0.950 & 0.867 & 0.887 & 0.015 \\
SVSNet+\_PF(R) & X & X & X & O & X & O & LL & Regression & 0.497 & 0.496 & 1.010 & 0.951 & 0.929 & 0.005 \\
SVSNet+\_PW(R) & X & X & X & O & X & X & WS & Regression & \textbf{0.578} & \textbf{0.577} & \textbf{0.846} & \textbf{0.960} & \textbf{0.969} & 0.009 \\
SVSNet+\_PFW(R) & X & X & X & O & X & O & WS & Regression & 0.554 & 0.554 & 1.062 & 0.946 & 0.949 & 0.129 \\
\bottomrule
\end{tabular}}
\end{center}
\label{table.1}
\vspace{-2.0em}
\end{table*}

\subsubsection{Training details}

All models were implemented using PyTorch (v2.0.1) in Python 3.10. Each model was trained using an NVIDIA GeForce RTX 3090 with 24GB RAM. All utterances were downsampled to a sampling rate of 16 kHz. The linear layer after SFM has a hidden size of 256, and the first linear layer in the prediction module has a hidden size of 128. The output size is 1 for regression tasks and 4 for classification tasks. The Adam optimizer \cite{Kingma2014AdamAM} was used to train the models with a learning rate of 1e-4. For regression tasks, we used the MSE loss for training, while for classification tasks, we used the cross entropy (CE) loss. Model parameters were initialized by the default method of PyTorch. We trained each model for 30 epochs on the VCC2018 training set with a batch size of 5,  and evaluated the performance of the model for each epoch using the VCC2018 test set. The model with the best system-level performance on the VCC2018 test set was selected and tested on the VCC2020 test set.

\subsection{Results}

\subsubsection{SVSNet+ with WavLM-Large}

In SVSNet \cite{9716822}, the input waveform is encoded by an encoder composed of SincNet, rSWC and BLSTM modules. To enhance SVSNet, we integrate WavLM-Large \cite{9814838} into it and verify the necessity of modules in the original encoder. The results corresponding to the best system-level performance for each combination on the VCC2018 test set are shown in Table 1. We trained two types of SVSNet models, regression and classification, as baselines, termed SVSNet(R) and SVSNet(C), respectively. There are discrepancies between the reproduced models and those in \cite{9716822}, which may be caused by software version and hyperparameter differences. Our proposed models are termed SVSNet+. For in-depth analysis, we implemented the following operations: (1) whether an additional linear layer for dimension adjustment is used; (2) whether WavLM-Large is fine-tuned for the speaker voice similarity assessment task; (3) whether the weighted sum (WS) of representations from all transformer encoder layers or only the representation from the last layer (LL) is used.

From Table 1, several observations can be drawn.
First, when integrated with WavLM-Large, SVSNet+ outperforms SVSNet in system-level evaluation in most configurations.
Second, the rSWC and BLSTM modules used in SVSNet do not bring notable benefits to SVSNet+ (SVSNet+\_rBPL(R) vs. SVSNet+\_BPL(R) and SVSNet+\_BPL(R) vs. SVSNet+\_PL(R)). The reason may be that the transformer encoder in WavLM-Large is already good enough at capturing contextual information in the waveform.
Third, fine-tuning WavLM-Large during SVSNet+ training does not help (SVSNet+\_PLFW(R) vs. SVSNet+\_PLW(R) and SVSNet+\_PFW(R) vs. SVSNet+\_PW(R)). This may be due to the inappropriateness of fine-tuning large pre-trained models with small datasets. Such phenomenon was also mentioned in \cite{tamm22_interspeech}.
Fourth, utilizing the weighted-sum representation of WavLM-Large can improve performance compared to using the last-layer representation (SVSNet+\_PW(R) vs. SVSNet+\_P(R)).
Lastly, although the additional linear layer may not provide significant advantages to SVSNet+ (SVSNet+\_PLW(R) vs. SVSNet+\_PW(R)), it is still valuable for representation resizing, allowing a fairer comparison between SVSNet+ and SVSNet.
Among all models, the two best-performing models are SVSNet+\_PLW(R) and SVSNet+\_PW(R), which are regression types with weighted sum and no fine-tuning. Therefore, these two SVSNet+ configurations will be used in subsequent experiments.

\subsubsection{SVSNet+ with different SFMs}

Next, we compare the performance of SVSNet+ with different SFMs, including WavLM-Large, wav2vec 2.0-Large, HuBERT-Large, HuBERT-XLarge, MMS-300M, MMS-1B, Whisper-Medium, and Whisper-Large. The results for the models with and without an additional linear layer are shown in Tables 2 and 3, respectively. The number of transformer encoder layers, embedding dimension, and number of attention heads are noted after each model. From Table 2, it can be found that SVSNet+\_HuBERT-Large achieves the best performance in system-level LCC and SRCC, followed by SVSNet+\_Whisper-Large. In terms of system-level MSE, all models show excellent and almost equivalent performance, although SVSNet+\_wav2vec 2.0-Large performs slightly better than the others. No matter which SFM is employed, the proposed SVSNet+ consistently outperforms the original SVSNet in all system-level metrics. The results reconfirm the advantage of integrating pre-trained SFMs to extract speech representations. Furthermore, it's worth noting that larger SFMs do not always result in better performance. This may be attributed to the difference in characteristics between the training speech of the upstream model and the test speech of the downstream task. 

Comparing Table 3 with Table 2, it can be seen that removing the additional linear layer results in poorer performance in system-level metrics for most models. Only SVSNet+\_WavLM-Large slightly improves over its counterpart with the additional linear layer. Since WavLM jointly learned masked speech prediction and denoising on mixed audio during pre-training, the additional linear layer may blur the extracted speech features, potentially harming the performance of downstream tasks.

\begin{table}[htb]
\begin{center}
\caption{Performance of SVSNet+ with different SFMs (with the additional linear layer) evaluated on VCC2018.}
\vspace{-0.2cm}
\setlength{\tabcolsep}{1.0pt} 
\scalebox{0.65}{
\begin{tabular}{lcccccc}
\toprule
\multirow{2}{*}{\textbf{Method}} & \multicolumn{3}{c}{\textbf{Utterance-level}} & \multicolumn{3}{c}{\textbf{System-level}} \\ \cmidrule(r){2-4} \cmidrule(r){5-7}
& \textbf{LCC ↑} & \textbf{SRCC ↑} & \textbf{MSE ↓} & \textbf{LCC ↑} & \textbf{SRCC ↑} & \textbf{MSE ↓} \\
\midrule
SVSNet(R) & 0.530 & 0.531 & 0.952 & 0.924 & 0.902 & 0.009 \\
SVSNet+\_WavLM-Large [24/1024/12] & 0.567 & 0.567 & 0.879 & 0.953 & 0.957 & 0.005 \\
SVSNet+\_wav2vec 2.0-Large [24/1024/16] & 0.573 & 0.573 & 0.856 & 0.958 & 0.943 & \textbf{0.004} \\
SVSNet+\_HuBERT-Large [24/1024/16] & 0.569 & 0.570 & 0.866 & \textbf{0.967} & \textbf{0.960} & 0.006 \\
SVSNet+\_HuBERT-XLarge [48/1280/16] & \textbf{0.581} & \textbf{0.578} & \textbf{0.842} & 0.953 &  0.955 & 0.005 \\
SVSNet+\_MMS-300M [24/1024/16] & 0.524 & 0.525 & 0.959 & 0.954 & 0.943 & 0.006 \\
SVSNet+\_MMS-1B [48/1280/16] & 0.575 & 0.573 & 0.843 & 0.950 & 0.958 & 0.006 \\
SVSNet+\_Whisper-Medium [24/1024/16] & 0.506 & 0.512 & 1.022 & 0.945 & 0.941 & 0.007 \\
SVSNet+\_Whisper-Large [32/1280/20] & 0.561 & 0.564 & 0.876 & 0.961 & 0.956 & 0.006 \\
\bottomrule
\end{tabular}}
\end{center}
\label{table.2}
\vspace{-2.0em}
\end{table}

\begin{table}[htb]
\vspace{-0.5em}
\begin{center}
\caption{Performance of SVSNet+ with different SFMs (without the additional linear layer) evaluated on VCC2018.}
\vspace{-0.2cm}
\setlength{\tabcolsep}{1.0pt} 
\scalebox{0.65}{
\begin{tabular}{lcccccc}
\toprule
\multirow{2}{*}{\textbf{Method}} & \multicolumn{3}{c}{\textbf{Utterance-level}} & \multicolumn{3}{c}{\textbf{System-level}} \\ \cmidrule(r){2-4} \cmidrule(r){5-7}
& \textbf{LCC ↑} & \textbf{SRCC ↑} & \textbf{MSE ↓} & \textbf{LCC ↑} & \textbf{SRCC ↑} & \textbf{MSE ↓} \\
\midrule
SVSNet(R) & 0.530 & 0.531 & 0.952 & 0.924 & 0.902 & 0.009 \\
SVSNet+\_WavLM-Large [24/1024/12] & \textbf{0.578} & \textbf{0.577} & \textbf{0.846} & 0.960 & \textbf{0.969} & 0.009 \\
SVSNet+\_wav2vec 2.0-Large [24/1024/16] & 0.572 & 0.571 & 0.848 & 0.926 & 0.902 & 0.008 \\
SVSNet+\_HuBERT-Large [24/1024/16] & 0.566 & 0.564 & 0.866 & \textbf{0.964} & 0.959 & \textbf{0.004} \\
SVSNet+\_HuBERT-XLarge [48/1280/16] & 0.563 & 0.562 & 0.864 & 0.939 & 0.931 & 0.007 \\
SVSNet+\_MMS-300M [24/1024/16] & 0.560 & 0.555 & 0.869 & 0.943 & 0.941 & 0.006 \\
SVSNet+\_MMS-1B [48/1280/16] & 0.561 & 0.558 & 0.879 & 0.939 & 0.929 & 0.007 \\
SVSNet+\_Whisper-Medium [24/1024/16] & 0.554 & 0.553 & 0.892 & 0.951 & 0.949 & 0.013 \\
SVSNet+\_Whisper-Large [32/1280/20] & 0.553 & 0.553 & 0.906 & 0.956 & 0.938 & 0.006 \\
\bottomrule
\end{tabular}}
\end{center}
\label{table.3}
\vspace{-2.0em}
\end{table}

\subsubsection{Evaluated on VCC2020}

To evaluate the generalization ability of SVSNet+, the above models trained on VCC2018 were tested on the VCC2020 test set. The system-level evaluation results are shown in Table 4. Unlike VCC2018, VCC2020 consists of two tasks: intra-lingual semi-parallel VC and cross-lingual VC. Moreover, in VCC2018, most VC systems employed conventional vocoders, while in VCC2020, neural vocoders were more common. All of the above differences can lead to serious corpus mismatch.
Comparing Table 4 with Tables 2 and 3, due to corpus mismatch, the scores of all models on VCC2020 are worse than those reported on VCC2018. However, certain SVSNet+ models, such as SVSNet+\_HuBERT-Large and SVSNet+\_Whisper-Large, achieve fairly good performance on VCC2020, although there is still room for further improvement.

Examining the performance of SVSNet+ models with the additional linear layer in Table 4, we can see that almost all SVSNet+ models outperform SVSNet in all metrics except SVSNet+\_Whisper-Medium in MSE (1.175). 
The SVSNet+\_HuBERT-Large and SVSNet+\_Whisper-Large models with the best performance on VC2018 in Table 2 also achieve relatively high performance on VCC2020.
Surprisingly, SVSNet+\_wav2vec 2.0-Large performs very well and achieves the highest SRCC (0.910), although it does not perform particularly well compared to other models on VCC2018.

For SVSNet+ models without the additional linear layer, while most models outperform SVSNet, SVSNet+\_wav2vec 2.0-Large performs poorly in both LCC and SRCC. Without further fine-tuning the wav2vec 2.0 representation using the additional linear layer, the resulting SVSNet+ model generalizes poorly. In contrast, removing the additional linear layer benefits SVSNet+\_MMS-1B, which achieves the highest scores in LCC and SRCC among all models with the same settings. Since MMS-1B was pre-trained using a larger and more diverse set of data, it excels at extracting more intricate patterns. Additional processing by the linear layer may be detrimental to the extracted features.

\begin{table}[htb]
\begin{center}
\caption{Performance of SVSNet+ with different SFMs evaluated on VCC2020. Each model is the best-performing checkpoint selected from the VCC2018 system-level results.}
\vspace{-0.2cm}
\setlength{\tabcolsep}{1.0pt} 
\scalebox{0.65}{
\begin{tabular}{lcccccc}
\toprule
\multirow{3}{*}{\textbf{Method}} & \multicolumn{6}{c}{\textbf{System-level}} \\ \cline{2-7} & \multicolumn{3}{c|}{\textbf{w/ linear layer}} & \multicolumn{3}{c}{\textbf{w/o linear layer}} \\ \cline{2-7} & \textbf{LCC ↑} & \textbf{SRCC ↑} & \multicolumn{1}{c|}{\textbf{MSE ↓}} & \textbf{LCC ↑} & \textbf{SRCC ↑} & \textbf{MSE ↓} \\
\hline
SVSNet(R) & 0.745 & 0.713 & \multicolumn{1}{c|}{0.764} & 0.745 & 0.713 & 0.764 \\
SVSNet+\_WavLM-Large [24/1024/12] & 0.768 & 0.766 & \multicolumn{1}{c|}{0.535} & 0.822 & 0.851 & 0.402 \\
SVSNet+\_wav2vec 2.0-Large [24/1024/16] & 0.885 & \textbf{0.910} & \multicolumn{1}{c|}{0.502} & 0.694 & 0.630 & 0.526 \\
SVSNet+\_HuBERT-Large [24/1024/16] & 0.854 & 0.844 & \multicolumn{1}{c|}{0.487} & 0.831 & 0.806 & 0.227 \\
SVSNet+\_HuBERT-XLarge [48/1280/16] & 0.848 & 0.840 & \multicolumn{1}{c|}{0.499} & 0.760 & 0.746 & 0.395 \\
SVSNet+\_MMS-300M [24/1024/16] & 0.803 & 0.775 & \multicolumn{1}{c|}{0.571} & 0.770 & 0.732 & 0.486 \\
SVSNet+\_MMS-1B [48/1280/16] & 0.794 & 0.764 & \multicolumn{1}{c|}{\textbf{0.310}} & \textbf{0.865} & \textbf{0.852} & 0.339 \\
SVSNet+\_Whisper-Medium [24/1024/16] & 0.771 & 0.775 & \multicolumn{1}{c|}{1.175} & 0.838 & 0.808 & \textbf{0.224} \\
SVSNet+\_Whisper-Large [32/1280/20] & \textbf{0.892} & 0.886 & \multicolumn{1}{c|}{0.319} & 0.859 & 0.835 & 0.368 \\
\hline
\end{tabular}}
\end{center}
\label{table.4}
\vspace{-2.5em}
\end{table}

\section{Conclusions}

This study demonstrates that representations extracted by SFMs can effectively enhance the performance of speaker voice similarity assessment models. Experiments conducted on VCC2018 and VCC2020 show that SVSNet+ leveraging SFM surpasses its predecessor SVSNet. The results also show that for different SFMs, an additional linear layer can have significantly different effects on the performance of assessment models. Along this research path, in addition to integrating a single SFM into SVSNet+, we also conducted preliminary experiments on fused SFMs. By concatenating the representations extracted by HuBERT-Large and Whisper-Large as input, SVSNet+ can achieve better system-level performance on VCC2018, with LCC, SRCC, and MSE of 0.97, 0.969, and 0.004, respectively. We will conduct further research in this direction in the future.

\section{Acknowledgements}

This work was supported in part by the Co-creation Platform of the Speech-AI Research Center, Industry-Academia Innovation School, NYCU, under the framework of the National Key Fields Industry-University Cooperation and Skilled Personnel Training Act, from the Ministry of Education (MOE), the National Development Fund (NDF), and industry partners in Taiwan.

\bibliographystyle{IEEEtran}
\bibliography{mybib}

\end{document}